\documentclass[preprint,showpacs,preprintnumbers,amsmath,amssymb,superscriptaddress]{revtex4}

\usepackage{graphicx}
\usepackage{dcolumn}
\usepackage{bm}
\usepackage[dvips]{epsfig}

\newcommand{\unit}[1]{\,\mathrm{#1}}
\newcommand{\mean}[1]{\left\langle #1 \right\rangle}
\newcommand{\pd}[2]{\frac{\partial #1}{\partial #2}}
\newcommand{\ts}{t_\mathrm{s}}
\newcommand{\FP}{\mathcal L}
\newcommand{\kT}{\mathit{k_\mathrm{B}T}}

\begin{document}


\title{Thermodynamics of a Colloidal Particle in a Time-Dependent Non-Harmonic
  Potential}

\author{V. Blickle}
\affiliation{2. Physikalisches Institut, Universit\"at Stuttgart,
Pfaffenwaldring 57, 70550 Stuttgart, Germany}

\author{T. Speck}
\affiliation{{II.} Institut f\"ur Theoretische Physik,
Universit\"at Stuttgart, Pfaffenwaldring 57, 70550 Stuttgart,
Germany}

\author{L. Helden}
\affiliation{2. Physikalisches Institut, Universit\"at Stuttgart,
Pfaffenwaldring 57, 70550 Stuttgart, Germany}

\author{U. Seifert}
\affiliation{{II.} Institut f\"ur Theoretische Physik,
Universit\"at Stuttgart, Pfaffenwaldring 57, 70550 Stuttgart,
Germany}

\author{C. Bechinger}
\affiliation{2. Physikalisches Institut, Universit\"at Stuttgart,
Pfaffenwaldring 57, 70550 Stuttgart, Germany}

\begin{abstract}
  We study the motion of an overdamped colloidal particle in a time-dependent
  non-harmonic potential. We demonstrate the first law-like balance between
  applied work, exchanged heat, and internal energy on the level of a single
  trajectory. The observed distribution of applied work is distinctly
  non-Gaussian in good agreement with numerical calculations. Both the
  Jarzynski relation and a detailed fluctuation theorem are verified with good
  accuracy.
\end{abstract}

\maketitle

Since more than a century, the first law relating the work applied to a system
with both the exchanged heat and an increase in internal energy is one of the
corner-stones of macroscopic physics. Its consistent formulation for a
mesoscopic system like a driven colloidal particle, however, was suggested
only about a decade ago~\cite{seki98}. Since on these scales thermal
fluctuations are relevant, probability distributions for work, heat and
internal energy replace the sharp values of their macroscopic counterparts.
Various theoretical relations like the fluctuation
theorem~\cite{evan93,gall95}, the Jarzynski relation~\cite{jarz97}, and the
Hatano-Sasa relation~\cite{hata01} involving these distributions in different
settings extend the second law to the mesoscopic realm at least as long as the
notion of a constant temperature of the ambient heat bath remains meaningful
(for a review, see~\cite{bust05}).  Such theorems have been tested
experimentally using both biomolecules manipulated
mechanically~\cite{liph02,coll05} as well as colloidal particles in
time-dependent laser traps~\cite{wang02,carb04,trep04}. Common to all
colloidal experiments, so far, is that these laser traps generate a {\it
  harmonic} potential albeit with a time-dependent center or ``spring
constant''. Consequently, often the interesting distributions are Gaussian
even though for certain quantities non-Gaussian distributions can
occur~\cite{carb04,zon03a}.

In this Letter, we study the thermodynamics of single colloidal
trajectories in a time-dependent {\it non-harmonic} potential
which, generically, gives rise to non-Gaussian distributions. Only
for very short or very long trajectories, one expects Gaussian
distributions even in this non-harmonic case~\cite{spec04}. In
particular, we identify applied work, exchanged heat and change in
internal energy along a single trajectory and thus test the
consistency of these notions on this level, or, put differently,
illustrate the validity of the first law. We measure the
distribution of work in the non-Gaussian regime and compare it to
theoretical prediction. Such a comparison does not involve a
single fit parameter since all quantities are measured
independently, which is another advantage of colloidal systems.
Finally, we test the Jarzynski relation which expresses the free
energy difference between two equilibrium states in terms of the
nonequilibrium work spent in the transition between the two
states. Such an illustration of the Jarzynski relation in the
non-Gaussian regime comes timely given ongoing theoretical
criticism of its validity~\cite{cohe04,jarz04}.

In our study, particle trajectories were determined using total
internal reflection microscopy (TIRM), where a single colloidal
particle is illuminated under evanescent field conditions. This
field is created by total internal reflection of a laser beam at a
glass--water interface. The scattered intensity of a bead near the
interface is proportional to $\exp(-\zeta\, z)$, with $\zeta^{-1}$
the decay length of the evanescent field and $z$ the
particle--wall distance~\cite{prie99}. Measuring the scattered
intensity of a fluctuating Brownian particle as a function of time
thus yields its vertical position with a spatial resolution of
about $5\unit{nm}$.

We used an aqueous suspension of highly charged polystyrene beads
with radius $R = 2\unit{\mu m}$, which were illuminated with light
of wavelength $\lambda = 658.5\unit{nm}$. The particle
concentration was sufficiently low to guarantee that there was
only a single particle within the field of view. The penetration
depth was adjusted to $\zeta^{-1}\simeq 200\unit{nm}$ and the
scattered intensity was monitored with a photomultiplier at a data
acquisition rate of $\nu=2\unit{kHz}$. An additional focused laser
beam ($\lambda=1064\unit{nm}$, power $P\simeq 2\unit{mW}$) was
directed vertically from the top, which confined the particle
motion to an one-dimensional trajectory in $z$-direction.

\begin{figure}
  \includegraphics[width=8cm]{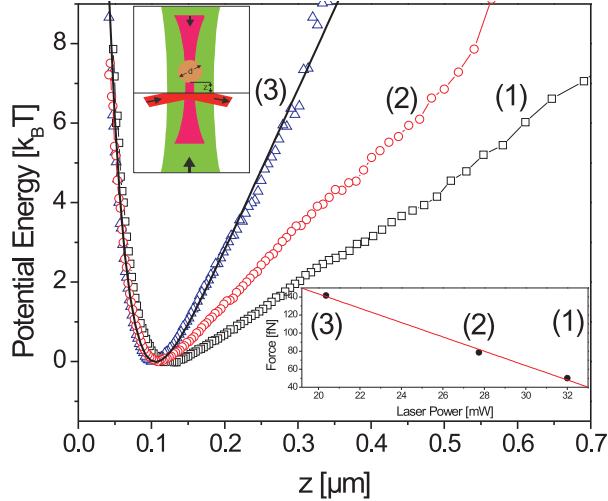}
  \caption{(color online) Particle wall interaction potentials for three different
    intensities of the lower optical tweezers [decreasing power from (1) to
    (3)]. The solid
    line shows the fit according to Eq.~\eqref{equ:pot} with
    $\kappa^{-1}\approx 25\unit{nm}$.  Inset: Light pressure versus tweezers
    intensity. The light pressure is a linear function of the laser
    intensity.}
  \label{fig:1}
\end{figure}

\begin{figure}
  \includegraphics[width=8cm]{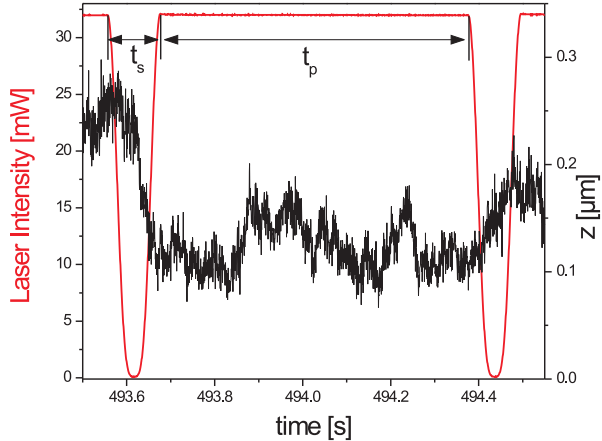}
  \caption{(color online) Measured tweezers intensity and particle trajectory. During the
    first pulse the particle is pressed towards the surface. During the second
    pulse thermal fluctuations support the particle and it is able to move
    away from the wall. Hence the applied work is positive for the first pulse
    and negative for the second.}
  \label{fig:2}
\end{figure}

To drive the colloidal particle between two equilibrium states, it
was subjected to the light pressure of another optical tweezers
($\lambda=532\unit{nm}$, beam waist about $17\unit{\mu m}$, $P\leq
60\unit{mW}$), which was incident from below into the sample cell
(lower tweezers, see inset of Fig.~\ref{fig:1}). The intensity of
this laser beam was varied with an electro-optical modulator
connected to a computer-controlled waveform generator. We
modulated the laser intensity according to a time-dependent
symmetric protocol $I(\tau)=I(\ts-\tau)$ in the time interval
$0\leq\tau\leq\ts$, where $\ts$ is the pulse duration (see Fig.
\ref{fig:2})~\footnote{Due to the characteristics of the
electrooptical modulator the pulse can be parameterized as
$I(\tau)\propto \sin\left[A_1\cos(2\pi\tau/\ts)+\varphi\right]$.}.
To ensure that the system is out of equilibrium, $\ts$ must be
smaller than the particle relaxation time $t_\mathrm{r}\simeq
480\unit{ms}$. On the other hand, when repeating the experiment
the pause $t_\mathrm{p}$ between two consecutive pulses must be
longer than $t_\mathrm{r}$ to guarantee equilibration of the
system. To meet both conditions in our experiments we have chosen
$t_\mathrm{s}=120\unit{ms}$ and $t_\mathrm{p}=700\unit{ms}$.

The total time-dependent potential acting on the particle at distance $z$ is
given by~\cite{prie99}
\begin{equation}
  V(z,\tau)=A_0\exp(-\kappa z)+B_{0}z+C_0I(\tau)z.
  \label{equ:pot}
\end{equation}
The first term describes the double-layer interaction between the
negatively charged colloidal particle and the likely charged wall
with $A_0$ depending on the corresponding surface charges and
$\kappa^{-1}$ the Debye screening length, which depends on the
salt concentration in the suspension. The second term accounts for
the weight of the particle and the additionally exerted light
pressure from the upper tweezers, which both depend linearly on
the particle distance $z$~\cite{blic05}~\footnote{The Rayleigh
lengths of the upper and the lower optical tweezers is $\simeq
4.6\unit{\mu m}$ and $425\unit{\mu m}$, respectively, therefore
they produce no force gradients in $z$-direction on the range of
the sampled particle positions.}. The last term considers the
time-dependent optical forces induced by the lower tweezers.
Experimentally, particle-wall potentials are easily obtained by
measuring the distance probability distribution $p(z)$ of a
colloid in front of a wall. In thermal equilibrium, i.e. for
$I(\tau)=\text{const.}$, the potentials are given by $V(z)=-\kT\ln
p(z)$ up to a constant. Here, $T$ is the temperature of the
environment and $\kT$ is the Boltzmann constant. The symbols in
Fig.~\ref{fig:1} show $V(z)$ obtained for three different
intensities of the lower tweezers. The solid line is a fit to
Eq.~\ref{equ:pot} (exemplarily shown only for one data set) and
clearly demonstrates that the particle is moving in a non harmonic
potential. It can be seen in Fig.~\ref{fig:1} that the light
pressure of the lower tweezers reduces the slope of the linear
part of the potential, as already demonstrated by other
authors~\cite{prie99, blic05}. The inset of Fig.~\ref{fig:1} shows
the expected linear dependence on the intensity of the light
pressure.

Fig.~\ref{fig:2} shows exemplarily the trajectory of a particle
driven by the time dependent potential~\eqref{equ:pot}. While
during the first pulse the particle is strongly forced towards the
surface, thermal fluctuations support it to move against the
applied force  away from the wall during the second pulse. This
clearly demonstrates that the particle is strongly coupled to the
surrounding heat bath.

In contrast to recent experimental
studies~\cite{liph02,coll05,wang02,carb04,trep04}, we want to test
experimentally to which precision energy conservation of the particle on its
trajectory is maintained. This does not only provide a rigorous check of the
experimental technique, data analysis, and the energy resolution but also
demonstrates the interplay of applied work and exchanged heat when the system
is non-adiabatically driven. Therefore, in addition to the work $W$ exerted on
the particle we need to determine its heat exchange $Q$ with the environment.

The Brownian motion of the colloidal bead is described by the Langevin
equation
\begin{equation}
  \gamma \dot{z}=-\frac{\partial V}{\partial z}+\xi,
\end{equation}
with $\gamma$ the friction coefficient and $\xi$ the stochastic force.
According to~\cite{seki98} the incremental change of heat $dQ$ and work $dW$ is
then given by
\begin{equation}
  dW=\frac{\partial V}{\partial \tau}d\tau, \quad
  dQ=-\frac{\partial V}{\partial z}dz.
  \label{equ:dqdw}
\end{equation}
Integration along a single trajectory $z(\tau)$ then leads to the work
functional
\begin{equation}
  W[z(\tau)]=\int^{\ts}_{0}d\tau\;\frac{\partial V}{\partial
    \tau}(z(\tau),\tau)
  = \frac{C_0}{\nu} \sum_{i}\dot{I}(\tau_i)z(\tau_i),
  \label{equ:w}
\end{equation}
where we have used Eq.~\eqref{equ:pot}. The right hand side of
Eq.~\eqref{equ:w} accounts for the discrete sampling of the particle
trajectory during our experiments with rate $\nu=\frac{1}{\delta t}$ at times
$\tau_i=i\delta t$. Because the velocity autocorrelation of a Brownian
particle decays on a timescale of some $10\unit{ns}$, the velocity
\begin{equation}
  \bar{\dot{z}}(\tau_i)=\nu\int^{\tau_{i+1}}_{\tau_i}d\tau\;\dot{z}(\tau)
\end{equation}
determined from a trajectory measured with $\nu = 2\unit{kHz}$ is not
identical to the instant particle velocity $\dot{z}$. However, since $\partial
V/\partial z$ varies on a timescale much larger than $\delta t$, the heat
along a single trajectory $z(\tau)$ can be written as
\begin{equation}
\begin{split}
  Q[z(\tau)] &=-\int_{0}^{\ts}d\tau\;\pd{V}{z}(z(\tau),\tau)\dot{z}(\tau) \\
  &=-\frac{1}{\nu}\sum_{i}\pd{V}{z}(z(\tau_i),\tau_i)\bar{\dot{z}}(\tau_i).
\end{split}
\end{equation}
With the above sign convention the heat $Q$ is negative (positive) when
extracted (delivered) from (to) the thermal environment. Since we have full
knowledge of the time-dependence of $I$, $V$, and $z$, both quantities $W$ and
$Q$ can be determined from our experiments.

\begin{figure}
  \includegraphics[width=8cm]{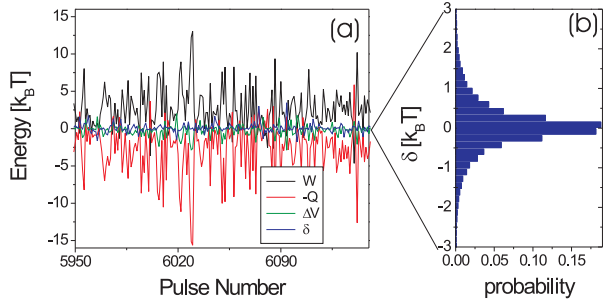}
  \caption{(color online) (a) The quantities $-Q$, $W$ and $\Delta V$ for about 100 periods
    of the protocol $I(\tau)$. (b) Distribution histogram of
    $\delta=W-Q-\Delta V$, the experimentally observed ``deviation'' from the
    first law of thermodynamics.}
  \label{fig:3}
\end{figure}

Introducing $\Delta V =V(z(\ts),\ts)-V(z(0),0)\label{equ:deltav}$ we finally
obtain a stochastic version
\begin{equation}
  W[z(\tau)] - Q[z(\tau)] - \Delta V = 0
  \label{equ:energieerh}
\end{equation}
of the first law of thermodynamics. Fig.~\ref{fig:3}(a) shows work
$W$, heat $Q$ and change of inner energy $\Delta V$ for the
trajectory of a single particle where the protocol $I(\tau)$ was
repeated about 100 times. For $W$ and $Q$ maximal energies of
about $15\,\kT$ are observed, whereas $\Delta V$ is on the order
of a few $\kT$. Obviously, $Q$ and $W$ are not independent
quantities. Usually trajectories resulting in a large work $W$ are
also accompanied by a large value of $Q$. But only when taking all
three energies in Eq.~\eqref{equ:energieerh} into consideration,
the distribution of the deviation shown in Fig.~\ref{fig:3}(b) is
centered around zero, having a half-width of about $0.7\,\kT$.
Assuming that the three terms have the same contribution to the
total error, the energy error of these energies is about one
quarter of $\kT$.

\begin{figure}
  \includegraphics[width=8cm]{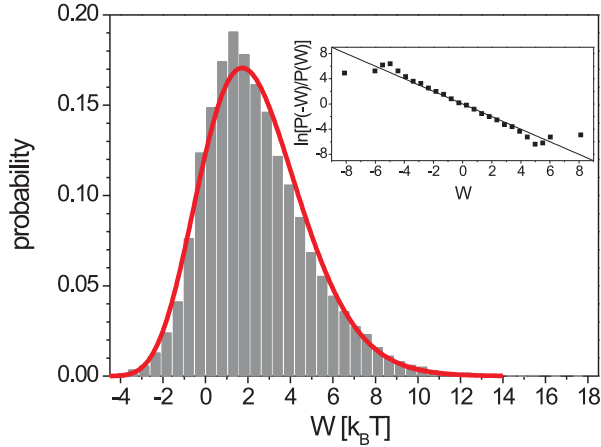}
  \caption{(color online) Non-Gaussian work distribution. The data was taken from about 16000
    trajectories, where the average work done on the particle was about $2.4\,
    \kT$. The solid line shows the Pearson type III
    distribution~\cite{pearson} corresponding to the theoretically
    obtained moments. Inset: Logarithm of the ratio of the probability to find
    trajectories with work $-W$ to those with work $+W$. The solid
    line shows the expected curve~\eqref{eq:ft}.  The deviation is due to the poor
    statistics of large negative work values $W\lesssim-4\,\kT$.}
  \label{fig:4}
\end{figure}

The measured work distribution in Fig.~\ref{fig:4} is distinctly
non-Gaussian and exhibits an asymmetry towards larger work values.
Whereas the first law is valid along a single trajectory as
demonstrated, fluctuation theorems considering probability
distributions can be regarded as an extension of the second law.
This becomes evident when looking at the Jarzynski
relation~\cite{jarz97}
\begin{equation}
  \mean{e^{-W/\kT}} = e^{-\Delta F/\kT},
  \label{eq:jarz}
\end{equation}
where $\Delta F$ is the change of free energy between two
equilibrium states and the brackets $\mean{\cdots}$ represent the
average over the work distribution spent in a transition between
these states. Eq.~\eqref{eq:jarz} immediately leads to
$\mean{W}\geq\Delta F$, a formulation of the second law for driven
systems on the mesoscopic scale.  A test of the Jarzynski relation
using the data shown in Fig.~\ref{fig:4} yields
$\mean{\exp(-W/\kT)}\simeq1.03$ in agreement with $\Delta F=0$ for
the symmetric protocol $I(\tau)$. In addition to the integral
theorem~\eqref{eq:jarz} we also test the somewhat stronger
detailed fluctuation theorem
\begin{equation}
  P(-W)/P(+W) = e^{-W/\kT},
  \label{eq:ft}
\end{equation}
which holds for time-symmetric protocols $I(\tau)=I(\ts-\tau)$
(see inset of Fig.~\ref{fig:4})~\cite{croo00,schu05}. Here the
probability $P(-W)$ that a negative work value occurs is compared
to the probability $P(+W)$ of a positive value of same magnitude.

\begin{table}
  \begin{tabular}{c|c|c|c}
    & $m_1$ [$\kT$] & $m_2$ [$(\kT)^2$] & $m_3$ [$(\kT)^3$] \\
    \hline
    Exp.  & 2.4 & 11.7 & 67.8 \\
    \hline
    Theo. & 2.4 & 11.6 & 63.7 \\
  \end{tabular}
  \caption{Comparison between theoretically predicted and measured moments of
    the work probability based on the data shown in Fig.~\ref{fig:4}.}
  \label{tab:1}
\end{table}

In order to compare the measured histogram in Fig.~\ref{fig:4} to the
theoretical prediction we calculate the probability distribution solving the
Fokker-Planck equation~\cite{spec04,impa05}
\begin{equation}
  \label{eq:fp}
  \pd{\rho}{t} = \FP\rho - \pd{V}{\tau}\pd{\rho}{w}.
\end{equation}
Here, $\rho(z,w,t)$ is the joint probability of the particle to be at time $t$
a distance $z$ away from the wall and to have accumulated an amount of work
$w$ up to this time. The Fokker-Planck operator~\cite{risken}
\begin{equation}
  \FP = \pd{}{z}\left(\frac{D_\perp}{\kT}\pd{V}{z} +
    \frac{3}{2}\pd{D_\perp}{z} + D_\perp\pd{}{z}\right)
\end{equation}
governs the dynamics of the particle, where
\begin{equation}
  D_\perp(z) \approx D_0[1+R/(z-R)]^{-1}
\end{equation}
is the diffusion coefficient for perpendicular motion near a
surface~\cite{bren61}. The free diffusion constant for a particle
with radius $R=2\unit{\mu m}$ is $D_0\simeq0.1\unit{\mu m^2/s}$ at
room temperature $T=293\unit{K}$. Since we start in equilibrium
with no work spent on the particle yet, the initial distribution
needed to solve Eq.~\eqref{eq:fp} is
\begin{equation}
  \rho(z,w,0) =
  \delta(w)\frac{\exp(-V(z,0)/\kT)}{\int dz\;\exp(-V(z,0)/\kT)}.
\end{equation}

Eq.~\eqref{eq:fp} is a Fokker-Planck equation in two space dimensions $z$ and
$w$ including a singular initial condition, which numerically is not easy to
handle. We therefore multiply Eq.~\eqref{eq:fp} with $w^n$ and integrate over
$w$. After one integration by parts we obtain the inhomogeneous evolution
equation
\begin{equation}
  \label{eq:mom}
  \pd{M_n}{t} = \FP M_n + n\pd{V}{\tau}M_{n-1}
\end{equation}
for the conditional $n$th moment of the work
\begin{equation}
  M_n(z,t) = \int_{-\infty}^{+\infty} dw\; w^n \rho(z,w,t).
\end{equation}
The actual $n$th moment $m_n$ then follows simply by integrating
over $z$ and thus adding the contributions of all possible final
positions of the particle. The function $M_0(z,t)$ is the
probability distribution of the position $z$ of the particle and
hence $m_0=1$. Table~\ref{tab:1} compares the numerically and
experimentally obtained first three moments of the data shown in
Fig.~\ref{fig:4}. We stress that this good agreement does not
involve a single fit parameter.

In summary, we have confirmed experimentally both a stochastic
formulation of the first law and various recent theoretical
ramifications of the second law in a time-dependent non-harmonic
potential, where the underlying distributions are typically
non-Gaussian. In the next step, non-harmonic systems with broken
detailed balance should be investigated to test
theorems~\cite{hata01,spec05a} which, so far, have been under
experimental scrutiny in the harmonic case only~\cite{trep04}.

\end{document}